\documentclass{aa}  

\usepackage{graphicx}
\usepackage{txfonts}
\usepackage{hyperref}

\hypersetup{
    colorlinks=false,
    linkcolor=blue,
    filecolor=magenta,      
    urlcolor=cyan,
    pdftitle={Cosmology Ruler Bookmark for Teaching and Outreach Purposes},
    pdfpagemode=FullScreen,
    }

\begin{document}

   \title{Cosmology Ruler Bookmark for Teaching and Outreach Purposes}

   \subtitle{Pen-and-pencil cosmological ruler calculator for everyone, especially students}

   \author{H. Dole
          \inst{1}
          }

   \institute{Universit\'e Paris-Saclay, CNRS, Institut d'Astrophysique Spatiale, 91405 Orsay, France\\
              \email{Herve.Dole@universite-paris-saclay.fr}
             }

   \date{Release: January 2024.}
   
\titlerunning{Cosmology Ruler Bookmark}


  \abstract
   {Cosmology in general, and relation between redshift and cosmic epoch in particular, is usually obscure to first years university students, secondary students, as well as journalists, politicians and the general public scientists may have interactions with. I identify the need for a simple artifact scientists may give to the public to clarify a few relations between redshift and other physical quantities, more meaningful for a non-scientist audience.}
   {This simple bookmark aims at completing previous "pen-and-pencil cosmological calculator" nomograms.}
   {I created a small, handy, duplicable bookmark with two printed sides, showing the corresponding cosmological values of redshift, age, time, and angular scale (for 1 kpc), using the Planck 2018 cosmology.}
   {On the {\it recto}, the redshift range of $[0.1, 1000]$ approaches the recombination with a logarithmic scale. On the {\it verso}, the redshift range is chosen to be $[0, 30]$ using a linear scale, covering the range of current (and future) detections of galaxies. A few examples are given, illustrating e.g. Planck, JWST or Euclid capabilities and complementarities, time interval non-linearity, properties of galaxies and clusters.}
   {This handy bookmark may be printed cheaply and offered to every student in physics (undergrad and grad student) in our universities or to secondary schools students we visit. The Cosmology Ruler Bookmark included is ready to print (single- or double-sided). The python script is available on github, allowing changes adapted to everyone's needs for teaching and outreach purposes, including with other cosmologies or applied to other scientific fields.}


   \maketitle
%

\section{Introduction \& Motivation}

   Astrophysicists and cosmologists sometimes interact professionally with people outside the scientific community. Teaching to first years university students, making an appearance in a nearby high- or middle-school, giving a public lecture, talking to journalists, politicians, or entrepreneurs (privately or publicly during events), chatting on social media, are among the many occasions to exchange with non-scientist populations, to promote the scientific approach, their research and discuss the latest results\footnote{These occasions of interacting with the public should be strengtened, as highlighted by e.g. UNESCO, which promotes science \& society relations in its recommandations for researchers, in particular "the need for science to meaningfully interact with society and vice-versa" \url{https://www.unesco.org/en/recommendation-science}}.
      
   Inevitably, the question of distances in the Universe rises. "How big is the Milky Way?", "How far is the galaxy you observed ? ", "What is the distance of the furthest galaxy detected ?", and many along this line are frequent questions asked. Expected answers usually involve distances in light-years or lookback time in Giga-year, but never in terms of redshift.

   As cosmologists mainly use the redshift as a way to manipulate the distances (in space and time), and because this quantity (and the attached concepts) is essentially unknown outside the scientific community, I found unfortunately useless and hopeless to try to explain the concept of redshift to a general audience. I experienced a few moments of pure loneliness in front of large audiences or with journalists when I tried to explain why a detection at redshift $z=2.5$ was super cool, but the time duration between $z=1$ and $z=2$ had nothing to compare with the time duration between $z=2$ and $z=3$. Not mentioning the inability to remember what lookback time corresponds to $z=4.5$.

   \cite{pilipenko2020} created a "paper-and-pencil cosmological calculator" that saved me frequently since then: I could easily read the correspondence between the redshift and the lookback time or the age of the Universe, given a reasonable cosmology in a matter of seconds when talking to a journalist or students. 

   Other tools relevant to cosmology and teaching or outreach include the essential Ned Wright's Cosmology Calculator{\footnote{\url{https://www.astro.ucla.edu/~wright/CosmoCalc.html}}} of \cite{wright2006}, or the map of the Universe by \cite{gott2005}, not mentioning the many variants of "the power of ten" movies online and, of course, many excellent books, articles, websites, smartphone apps, movies or interviews.

    There are three kinds of motivations to have a bookmark with the two sides printed with axes showing the relationships between redshift, age of the universe, lookback time and angular distance scale. The first is about the content:
    \begin{itemize}
        \item easy and quick reference tool for scientists
        \item handy when teaching \& working with university students, for theory or data analysis purposes or at the telescope
        \item allows to have a sense of orders of magnitudes
        \item relates the redshift with other observables and physical quantities
        \item shows the non-linearity of time intervals for a fixed redshift interval $\Delta z$
        \item can represent non-linear (e.g. time vs redshift) or non-monotonic (e.g. angular diameter distance) relations
    \end{itemize}
    
    The second is about the format:
    \begin{itemize}
        \item it has to be small, lightweight and handy, so we can have it everywhere, including in books or in the classroom
        \item no need to have an online access to use it (particularly useful for outreach)
        \item can be printed on single-sided printers (and can be folded to get the two sides) or double-sided printers
        \item the python code being available, the bookmark can be adapted to everyone's needs
        \item scalable and eco-friendly: it has to be cheap and as ecological as possible to produce and print, in the spirit of sustainability of research \citep[among others]{knodlseder2023}
    \end{itemize}

    The third about the target audience:
     \begin{itemize}
        \item university students, in a lecture, during a lab work or at the telescope
        \item secondary school students
        \item general public, including in large venues
        \item journalists or other non-scientists
        \item it can be used as a smart and frugal goodie
        \item it can be printed and offered to large audiences
        \item almost no cost, but high scientific content inside 
    \end{itemize}

\cite{pilipenko2020} paper is a major inspiration to create another analog cosmological calculator in a complementing format, as a bookmark. Another source of inspiration is the old-fashioned slide rule allowing e.g. fast calculations of logarithms or multiplications, that were still in use at the beginning of the space age, and based on the nomogram. 
Notice that despite its originality and strengths, the bookmark is, by design, less accurate than the other tools, given its compactness.

\section{The Cosmology Ruler Bookmark}

I use the \verb|Planck2018| cosmology from \cite{Planck2020I} in the \verb|astropy.cosmology| package \citep{astropy:2013, astropy:2018, astropy:2022}. The bookmark has 2 sides of 5 lines. Parameters are given in Tab.~\ref{Tab:parameters}.

The top and bottom plot the redshift $z$ scale, to facilitate the reading. 
On the second line is plotted the corresponding age in Gyr. The third line contains the lookback time in Gyr. The fourth line represents the angular diameter distance of 1 kpc proper at $z$, in arcsecond.

The {\it recto} side covers a redshift range between 0.1 and 1000 in logarithmic scale, where dark energy ($z < 0.7$) and matter dominates the expansion of the post-recombination Universe. 

The {\it verso} side covers redshifts ranging between 1 and 30 in linear scale, where most of extragalactic sources are (or will be soon) detected. The bottom redshift scale has more ticks, every $\Delta z = 0.5$. Notice the irregular placement of intermediate ticks for the age and lookback time, for a better reading accuracy.

The bookmark has an aspect ratio of 6:1. The nominal size of the printed bookmark is $24~\rm{cm} \times 4~\rm{cm}$ or $21~\rm{cm} \times 3.5~\rm{cm}$.

\begin{figure*}
   \centering
   \includegraphics[angle=0, height=21.5cm]{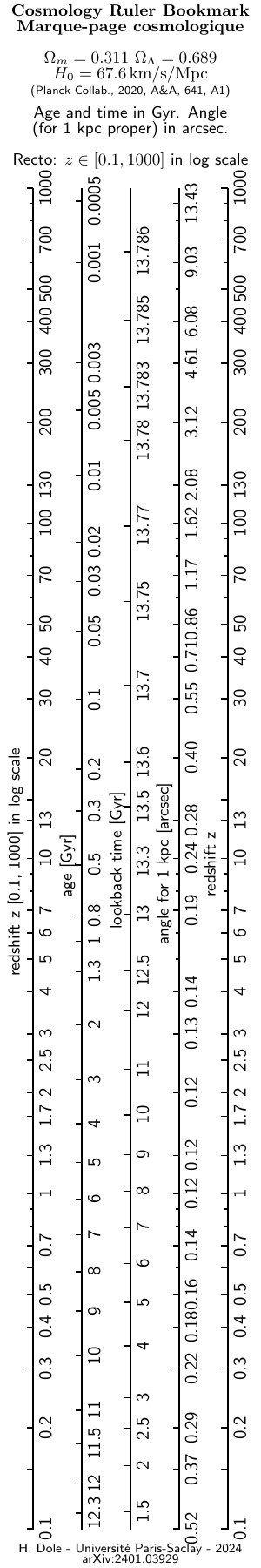}
   \includegraphics[angle=0, height=21.5cm]{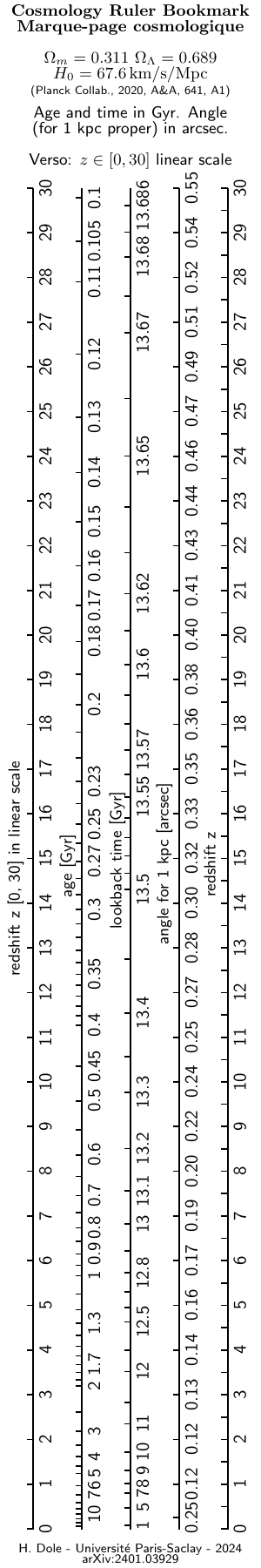}
   \includegraphics[angle=0, height=21.5cm]{CosmologyRulerBookmark_Recto.pdf}
   \includegraphics[angle=0, height=21.5cm]{CosmologyRulerBookmark_Verso.pdf}
      \caption{Two specimen of the two sides of the Cosmology Ruler Bookmark, ready to print and cut, with the actual size $21~\rm{cm} \times 3.5~\rm{cm}$ and an aspect ratio of 6:1. Advice 1 (if using a single-sided printer): print this page; cut the page vertically in the middle to separate the two bookmarks. Then fold each bookmark vertically in its middle to get two bookmarks. Advice 2 (if using a double-sided printer): print this page twice, double-sided on a single sheet of paper. Cut each of the four bookmarks.
     }
         \label{Fig:CRBrecto}
   \end{figure*}

The Cosmology Ruler Bookmark python script is available on github\footnote{\url{https://github.com/hrvdole/cosmology_ruler_bookmark}}. The bookmark is reproduced in Fig.~\ref{Fig:CRBrecto}.

\begin{table}[hb]
 \caption[]{Cosmological parameters from \cite{Planck2020I} (Planck+BAO) and redshift scales used in the bookmark.}
         \label{Tab:parameters}
\begin{tabular}{l l }        
            \hline
            \noalign{\smallskip}
            Cosmological Parameters:       &   \\
            \noalign{\smallskip}
            \hline
            \noalign{\smallskip}
            $H_0$            & 67.66~km/s/Mpc    \\
            $\Omega_M$        &   0.311    \\
            $\Omega_{\Lambda}$ &  0.689 \\
            \noalign{\smallskip}
            \hline
            \noalign{\smallskip}
            Redshift ranges:       &   \\
            \noalign{\smallskip}
            \hline
              \noalign{\smallskip}
            {\it Recto}   & $z \in [0.1, 1000]$ log scale\\
            {\it Verso}   & $z \in [0, 30]$ linear scale \\
           \noalign{\smallskip}
            \hline
            \noalign{\smallskip}
            Axes:       &   Line (top to bottom) \\
            \noalign{\smallskip}
            \hline
              \noalign{\smallskip}
            Redshift z  & 1 \& 5 \\
            age [Gyr] at $z$  & 2 \\
            lookback time [Gyr] at $z$& 3 \\
            angular scale for 1 kpc proper [arcsec] & 4 \\
            \noalign{\smallskip}
            \hline

\end{tabular}
\end{table}

\section{Examples of use}

\subsection{At the University with students} 

\cite{wang2023} report the redshift confirmation of a far away galaxy with JWST, at $z=12.39$, with a half-light radius of $\sim 0.17\arcsec$. Using the bookmark ({\it verso}: linear redshift scale between 0 and 30), students can search and find that the corresponding age is around 350~Myr after the Big Bang, and the lookback time around 13.45 Gyr ago. At $z \sim 12.4$, we read the angular diameter distance of $\sim 0.275 \arcsec /$kpc. Thus the measured $\sim 0.17\arcsec$ correspond to $\sim 620$pc (ignoring the lensing effects).

\cite{laporte2022} report the presence of a lensed galaxy protocluster at $z=7.66$ with JWST. Searching for the age (resp. lookback time), we read: 680~Myr after the Big Bang (resp. 13.1~Gyr ago).
The maximum angular separation of the galaxy members is about $30\arcsec$. Reading $\sim 0.197\arcsec /$kpc at this $z$, the physical size of this structure can thus be estimated around $150$kpc.

At the telescope, students detect in imaging the quasar QSO J0831+5245 at $z=3.91$ after a few minutes of integration. Still using the {\it verso} side of the bookmark with the linear redshift scale, we read the age of $\sim 1.55$~Gyr, and the lookback time of around $12.25$~Gyr. Incidentally, the redshift allows to estimate of the CMB temperature at this cosmic epoch (here $T_{CMB}(z=3.91) = (1+z) T_0 \sim 13.4$K) as well as the scale factor evolution $a_0/a$.

Planck \citep{Planck2016PSZ2} detected the massive galaxy cluster PLCK G266.6-27.3 at a redshift $z=0.972$ with an signal-to-noise ratio of 7.7 in the Sunyaev-Zeldovich (SZ) effect. We can round it at $z=1$. Assuming an angular diameter of $5\arcmin$ (the Planck beam), and reading the angular diameter of $0.12\arcsec$/kpc on the bookmark, we deduce an upper limit on the physical diameter of the cluster around 2.5~Mpc. \cite{williamson2011} detected this galaxy cluster with SPT at a higher angular resolution, where most of the SZ flux (thus the baryons) lie within $90\arcsec$, translating to a physical diameter of about 750~kpc.
This kind of galaxy cluster is one of the many measurements participating to the $S_8$ tension \citep{abdalla2022,perivolaropoulos2022}.

Planck \citep{Planck2020I} observed the Cosmic Microwave Background (CMB) at $z \sim 1090$, outside (but close to) the scale limit of the {\it recto}'s logarithmic scale. Close enough to derive an upper limit of the age of the Universe at this redshift: $< 500 000$ years. 

On a theoretical topic, we can also use the {\it recto} side. The signal of reionisation may start at $z=200$ \citep{zaroubi2013}. We can infer the CMB temperature at that very redshift, of about 540K. We can read the age, of $\sim 6$~Myr and the lookback time of slightly more than... 13.781~Gyr.

Looking at the time interval $\Delta t$ for a fixed redshift interval, e.g. $\Delta z =1$ is also interesting. What is $\Delta t$ at redshifts $z=2$, $z=7$ and $z=23$ ? Reading the age on the {\it verso} linear redshift scale, we find about $\Delta t (z=2) \sim 900$~Myr, $\Delta t (z=7) \sim 300$~Myr and $\Delta t (z=7) \sim 90$~Myr.\\

\subsection{For Outreach} 

Presenting ongoing space missions is a relevant and attractive way to present astrophysics to kids. Using the {\it verso} of the bookmark (the linear redshift scale, easier to use), we can search what is the lookback time corresponding to the aforementioned galaxy at $z=12.39$ when presenting the JWST \citep{gardner2023}. We can also look at an exiting supernova triply detected by lensing at $z=1.78$ \citep{frye2023,polletta2023} behind a massive galaxy cluster. At this redshift, the lookback time is around 3.5~Gyr. This kind of $z>1$ supernova may help tackling the problem of the $H_0$ tension \citep{abdalla2022,perivolaropoulos2022}.

Euclid \citep{laureijs2011} is starting the huge extragalactic sky survey with exquisite angular resolution. We can search what is the lookback time for most of the galaxies that will be used to derive the cosmological parameters (up to $z \sim 1$), around 8~Gyr ago. The acceleration of the expansion took place around $z \sim 0.7$ (if using $\ddot{a} = 0$, $a$ being the scale factor), around 7~Gyr ago. 

We can also reuse the case of Planck.
Such examples allow to explain the different goals of these space missions, their various cosmological probes targeting different cosmological era.

When dealing with the "Cosmic Noon", "Cosmic Dawn" and epoch of reionization (EoR, \cite{zaroubi2013}), we may have the audience find what is the lookback time of $z \sim 2$ and $z \sim 6$, respectively: around 10.5~Gyr and 12.8~Gyr. And finally compute the fractional lookback time of the Universe at these cosmic epochs, respectively $\sim 75\%$ and $\sim 93\%$.

\section{Conclusion and Perspectives} 

The redshift $z$ is a prime observed quantity, independent of the cosmology used, directly related to the scale factor. $z$ is widely used in astrophysics and cosmology, but not in non-scientific worlds. Converting $z$ into ages or distances involves the use of cosmological parameters. Here I use the \cite{Planck2020I} parameters to create a two-side bookmark. It is possible to choose another cosmology.

This bookmark may be printed cheaply, and is designed to be frugal. It may be offered to every student in physics (undergrad and grad student) in our universities or to secondary schools students. It can also be used while talking to journalists, politicians, or other non-scientists.

The concept of log scales (in time or space or energy) on a handy and cheap bookmark may also be adapted to other fields of science like particle physics, geophysics, planetary sciences etc., since nomograms can be adapted to various situations. The two sides allow to explore different scale ranges or to address two topics at once. The Cosmology Ruler Bookmark python script is available on github, allowing changes adapted to everyone's needs for teaching and outreach purposes.

\begin{acknowledgements}
      The author warmly thanks Dr Brenda Frye and Dr Mari Polletta for their suggestions and kind help, and Dr Ad\'elie Gorce for her suggestions and quick adaptation to the redshifted 21cm line. This work made use of Astropy\footnote{\url{http://www.astropy.org}} (and Cosmology): a community-developed core Python package and an ecosystem of tools and resources for astronomy \citep{astropy:2013, astropy:2018, astropy:2022}. 
      \end{acknowledgements}

\newpage 

%

%
   \begin{figure*}
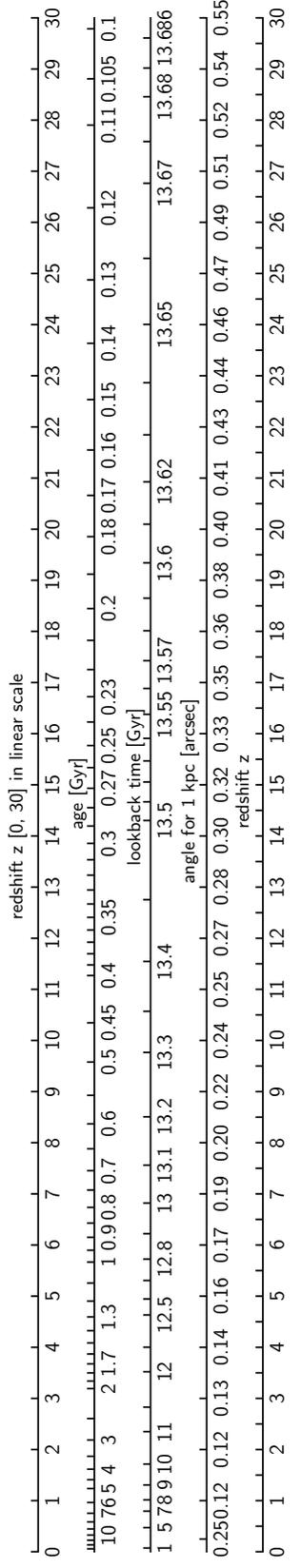

   \centering
   \includegraphics[angle=0, height=25cm]{CosmologyRulerBookmark_Recto.pdf}
   \includegraphics[angle=0, height=25cm]{CosmologyRulerBookmark_Verso.pdf}
   \includegraphics[angle=0, height=25cm]{CosmologyRulerBookmark_Recto.pdf}
   \includegraphics[angle=0, height=25cm]{CosmologyRulerBookmark_Verso.pdf}
      \caption{Same as Fig~\ref{Fig:CRBrecto}, but slightly bigger. 
     }
         \label{Fig:CRBrecto2}
   \end{figure*}


\end{document}